\documentclass[12pt,preprint]{emulateapj}
\usepackage{amssymb}
\usepackage{epsfig}

\newcommand{\etal}{et~al.~}

\journalinfo{ApJL, 8 Sep 2009, in press}

\begin{document}


\title{MMTF-H$\alpha$ and {\em HST}-FUV Imaging of the Filamentary
  Complex in Abell~1795}


\author{Michael McDonald\altaffilmark{1}, Sylvain
  Veilleux\altaffilmark{1,2}}

\altaffiltext{1}{Department of Astronomy, University of Maryland, College
  Park, MD 20742} 

\altaffiltext{2}{Also: Max-Planck-Institut f\"ur extraterrestrische
  Physik, Postfach 1312, D-85741 Garching, Germany}


\begin{abstract}

  We have obtained deep, high spatial resolution images of the central
  region of Abell~1795 at H$\alpha$ and [N~II] $\lambda$6583 with the
  Maryland Magellan Tunable Filter (MMTF), and in the far-ultraviolet
  (FUV) with the Advanced Camera for Surveys Solar Blind Channel on
  the {\em Hubble Space Telescope} ({\em HST}). The superb image
  quality of the MMTF data has made it possible to resolve the known
  SE filament into a pair of thin, intertwined filaments extending for
  $\sim$50 kpc, with a width $<$ 1 kpc. The presence of these thin,
  tangled strands is suggestive of a cooling wake where runaway
  cooling is taking place, perhaps aided by an enhanced magnetic field
  in this region.  The {\em HST} data further resolve these strands
  into chains of FUV-bright stellar clusters, indicating that these
  filaments are indeed sites of on-going star formation, but at a rate
  $\sim$2 orders of magnitude smaller than the mass-deposition rates
  predicted from the X-ray data. The elevated [N~II]/H$\alpha$ ratio
  and large spatial variations of the FUV/H$\alpha$ flux ratio across
  the filaments indicate that O-star photoionization is not solely
  responsible for the ionization. The data favor collisional heating
  by cosmic rays either produced in-situ by magnetohydrodynamical
  processes or conducted in from the surrounding intracluster medium.

\end{abstract}




\section{Introduction}

The absence of massive ($\sim 100 - 1000 $M$_{\odot}$ yr$^{-1}$ )
cooling flows in the cores of X-ray luminous galaxy clusters is often
used as prime evidence that feedback plays an important role in
regulating star formation and galaxy formation in dense environments
(see, e.g., review by Veilleux, Cecil, \& Bland-Hawthorn
2005). Energies of a few $\times10^{49}$ ergs per solar mass of stars
formed are needed to explain the sharp cutoff at the bright end of the
galaxy luminosity function. Starburst-driven winds are too feeble by a
factor of several to fully account for the cutoff, so AGN feedback is
invoked. The ubiquity of large ``cavities'' in the X-ray surface
brightness of clusters with radio galaxies confirms that AGN outflows
modify the thermodynamics of the intracluster medium (ICM; see review
by Peterson \& Fabian 2006). The relativistic gas injected into the
ICM by the AGN has enough energy to quench the mass accretion of
cooling flows, but the exact mechanism by which the energy in the
radio bubbles turns into heat is still debated.

The presence of warm hydrogen in the form of line-emitting filaments
extending from the brightest cluster galaxy (hereafter BCG) has been
observed in many cooling flow clusters to date (e.g. Hu \etal 1985,
Heckman \etal 1989, Crawford \etal 1999; Jaffe \etal 2005; Johnstone
\etal 2005). However, while several intriguing ideas have been put
forward, the origin of this gas and the mechanism for heating it are,
as yet, unknown. Of all the galaxy clusters with known optical
filaments, there is perhaps none quite so spectacular as
Abell~1795. This cluster has been very extensively studied at a
variety of wavelengths, leading to the discovery of a single, long
($\sim 50$ kpc) filament seen in H$\alpha$ (Cowie \etal 1983; Maloney
\& Bland-Hawthorn 2001; Jaffe \etal 2005) and X-ray (Fabian \etal
2001; Crawford \etal 2005), a powerful, double-sided, radio jet (4C
26.42; van Breugel \etal 1984, Ge \& Owen, 1993) emanating from the
central AGN, and a very disturbed, star-forming, central region
(McNamara \etal 1996; Smith \etal 1997; Mittaz \etal 2001; Crawford
\etal 2005).

As part of a survey of cooling flow clusters, we have carried
out deep, high-resolution (delivered image quality, DIQ $\sim
0.7\arcsec$) imaging of Abell~1795 at $H\alpha$ and [N~II]
$\lambda$6583 using the Maryland-Magellan Tunable Filter (MMTF) on the
Magellan-Baade 6.5-m telescope and in the far-ultraviolet (FUV) using
the ACS solar blind channel (SBC) camera on the {\em Hubble Space
  Telescope} ({\em HST}). These data far surpass any previously
available images of the filaments in this cluster in both depth and
spatial resolution; this {\em Letter} describes the results from our
analysis of these data. The results from the survey will be presented
in future papers.  The acquisition and reduction of the data on
Abell~1795 are discussed in Section 2, followed by a description of
the results (Section 3) and a discussion of the implications (Section
4). Throughout this {\em Letter} we assume a distance to Abell~1795 of
260 Mpc, based on a cosmology with $H_0=73$ km s$^{-1}$ Mpc$^{-1}$, $\Omega_{matter}$=0.27 and $\Omega_{vacuum}$=0.73.

\begin{figure*}[p]
\centering
\begin{tabular}{ccc}
\includegraphics[width=0.33\textwidth]{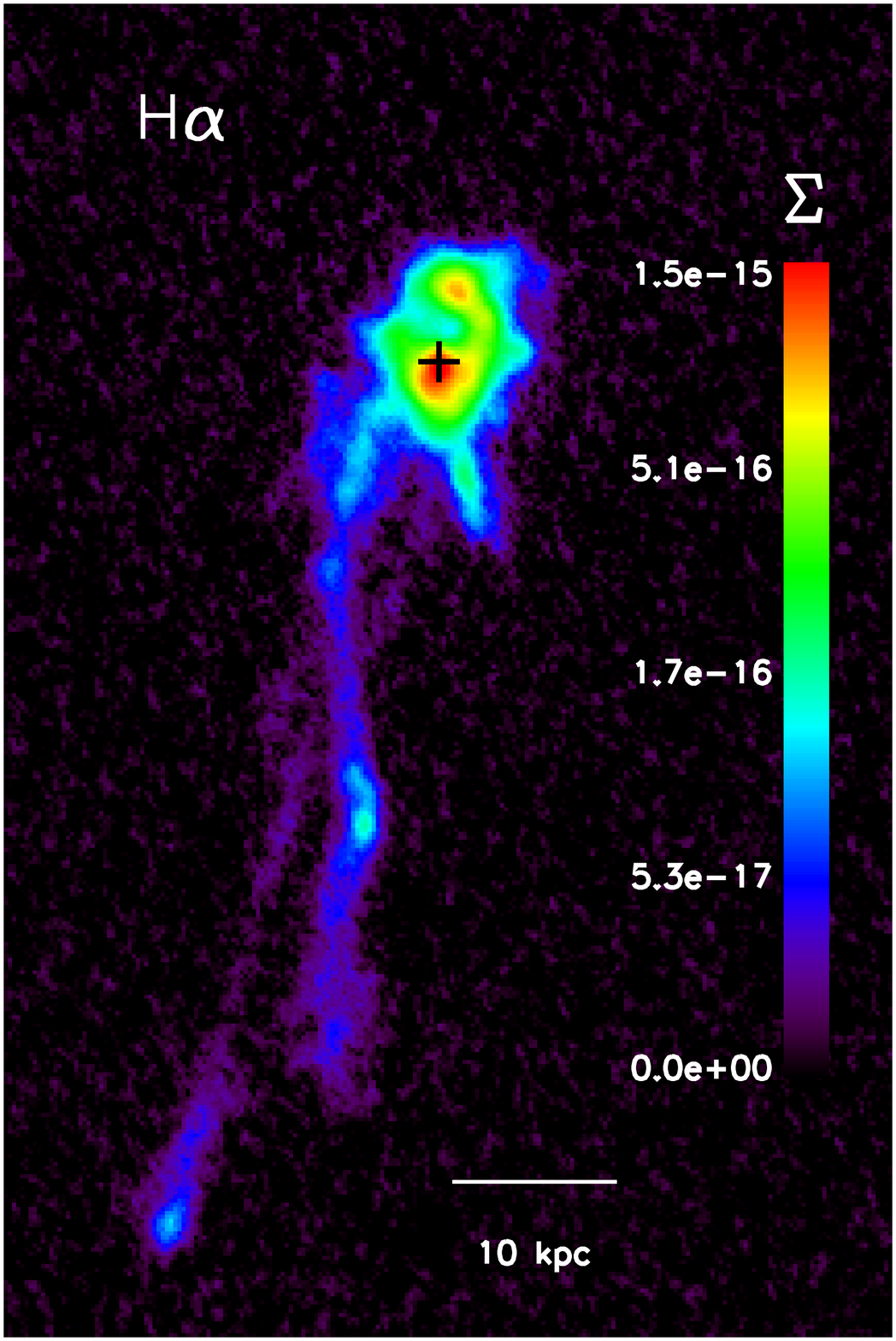} &
\includegraphics[width=0.33\textwidth]{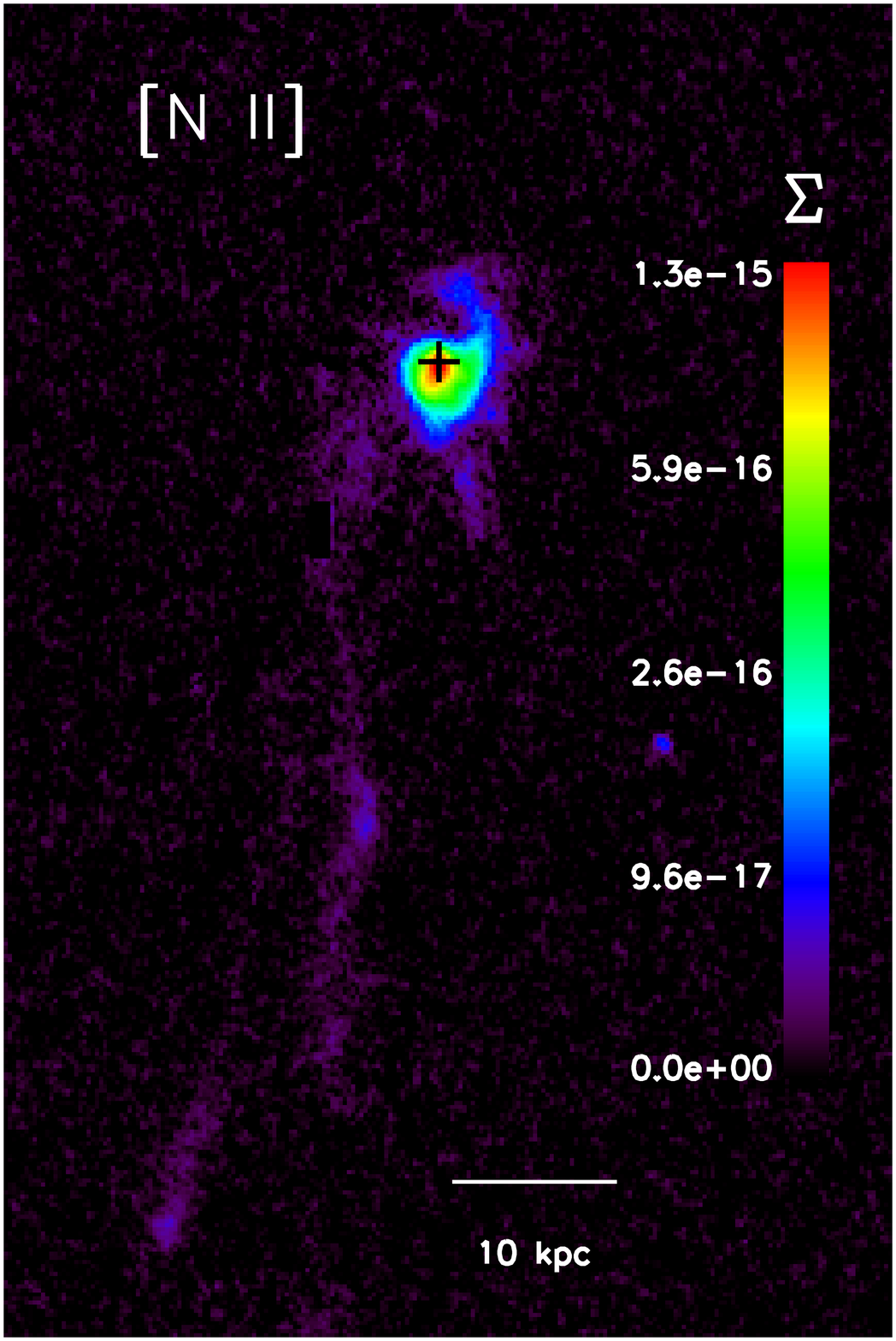} &
\includegraphics[width=0.33\textwidth]{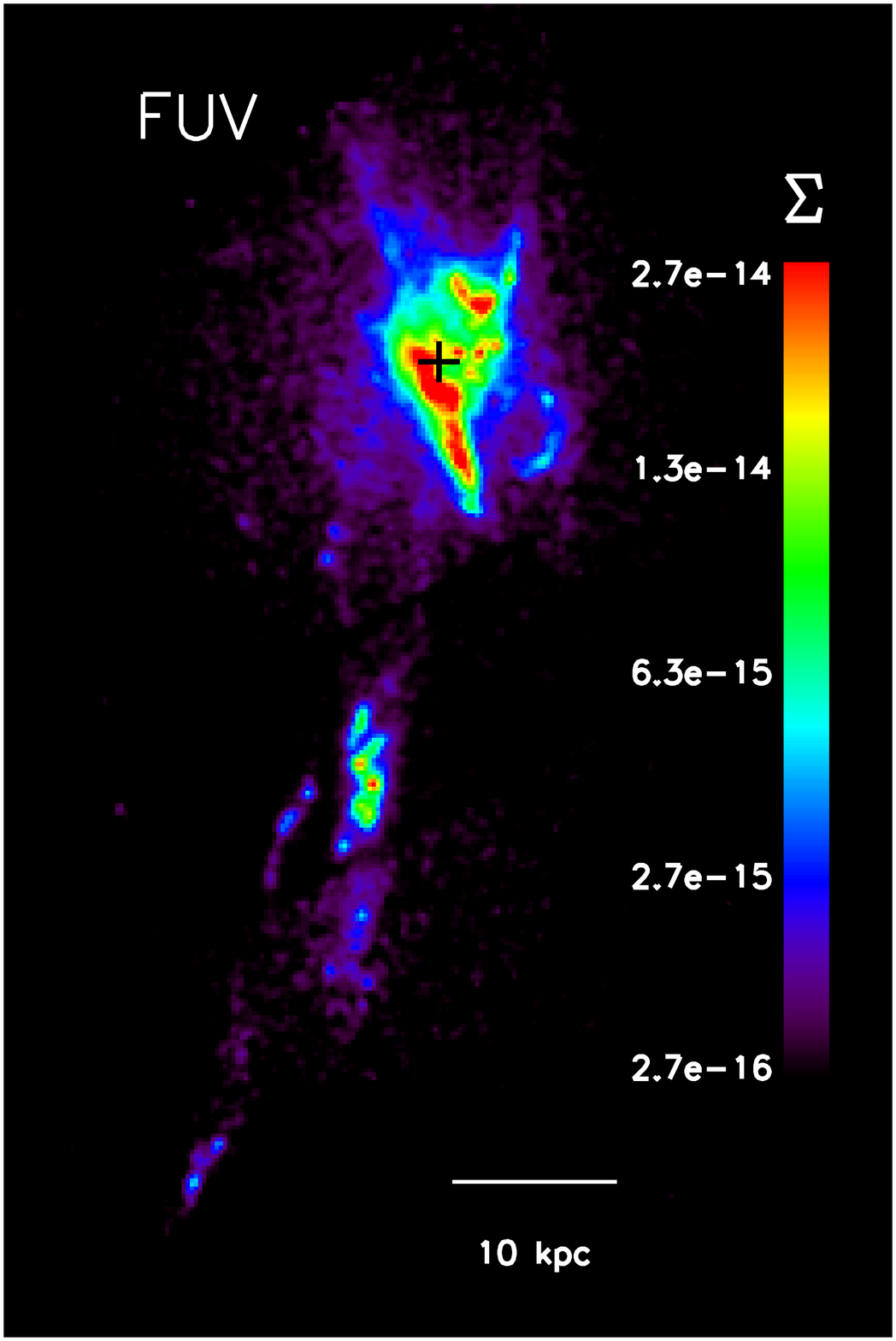} \\
\includegraphics[width=0.33\textwidth]{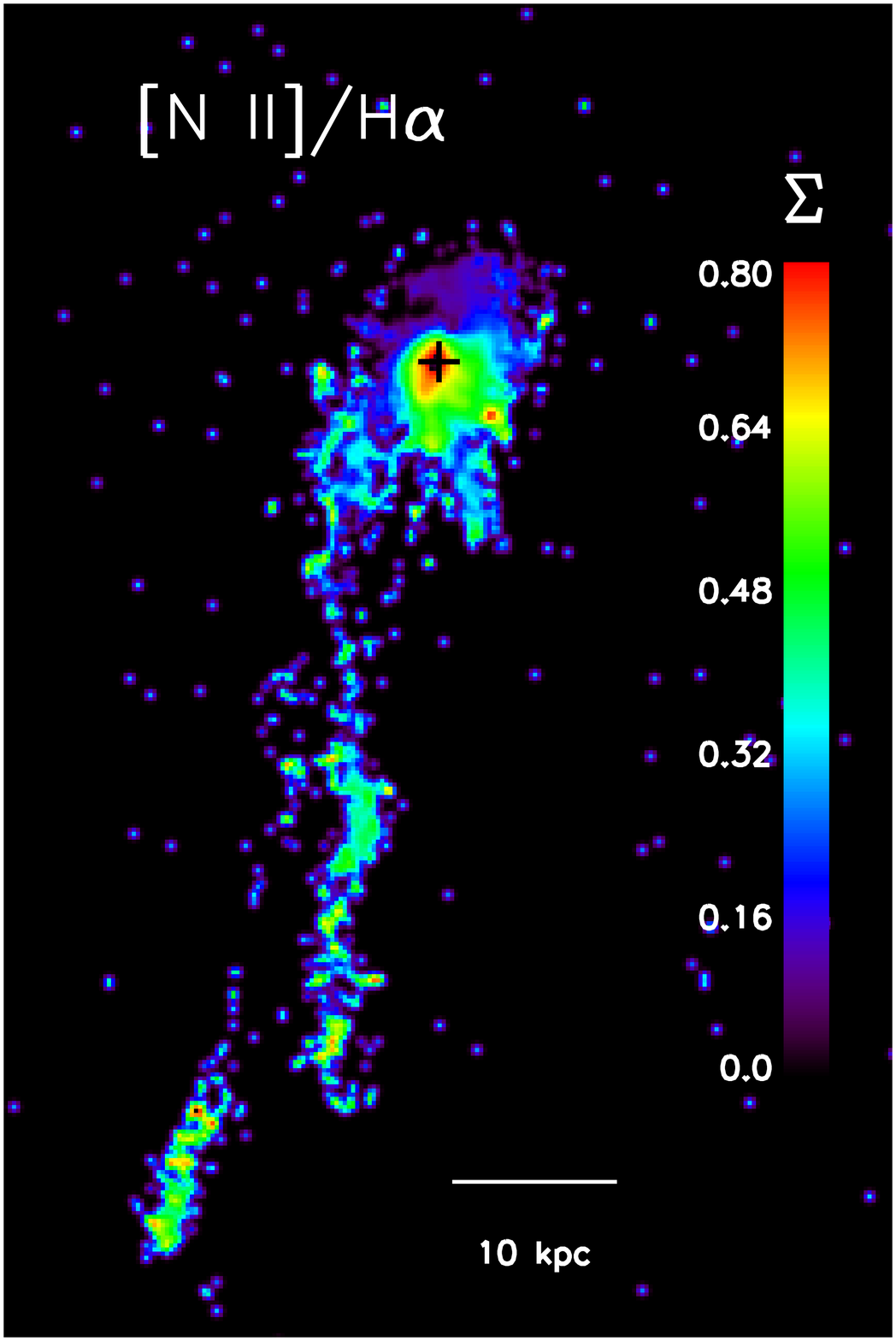} &
\includegraphics[width=0.33\textwidth]{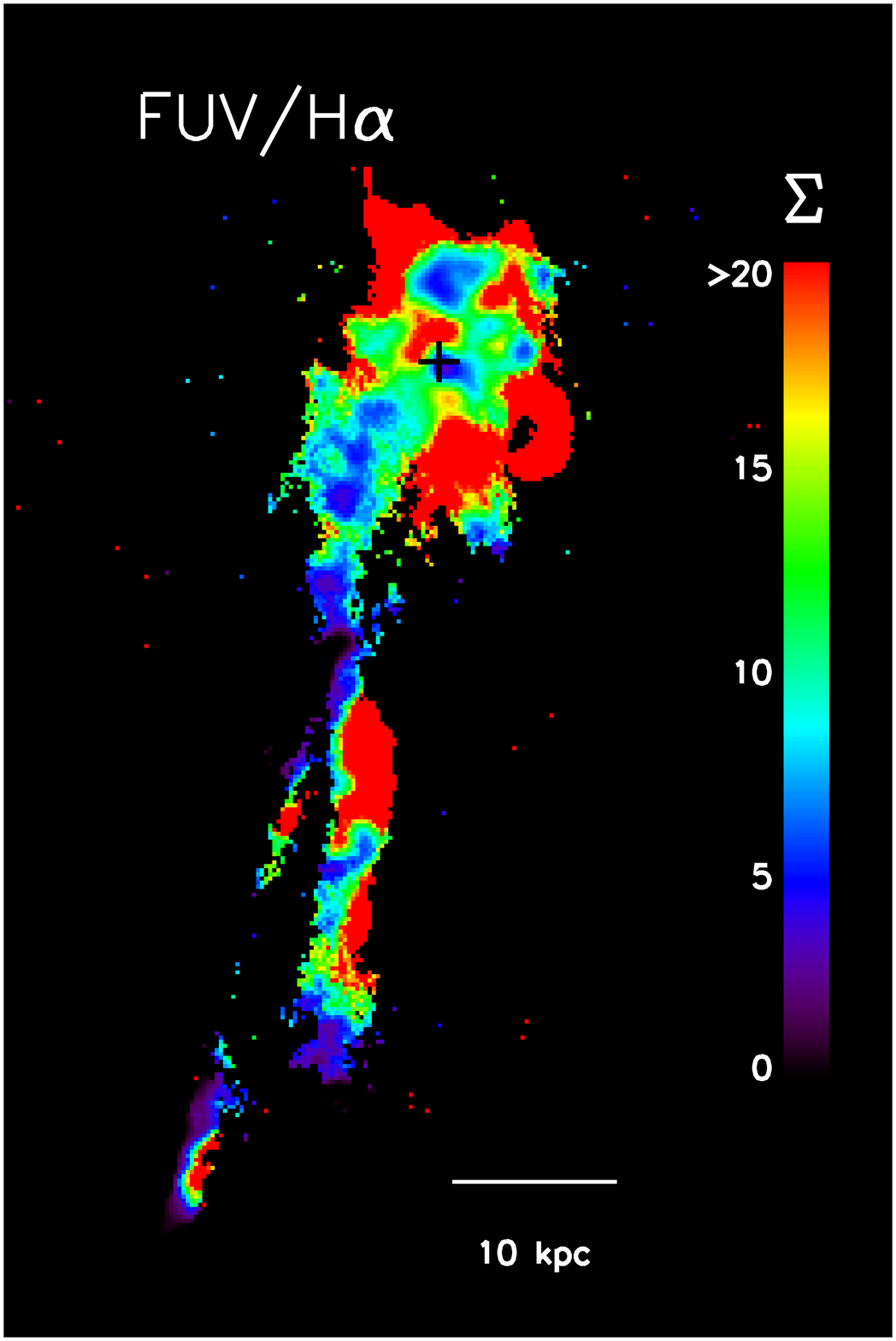} &
\includegraphics[width=0.33\textwidth]{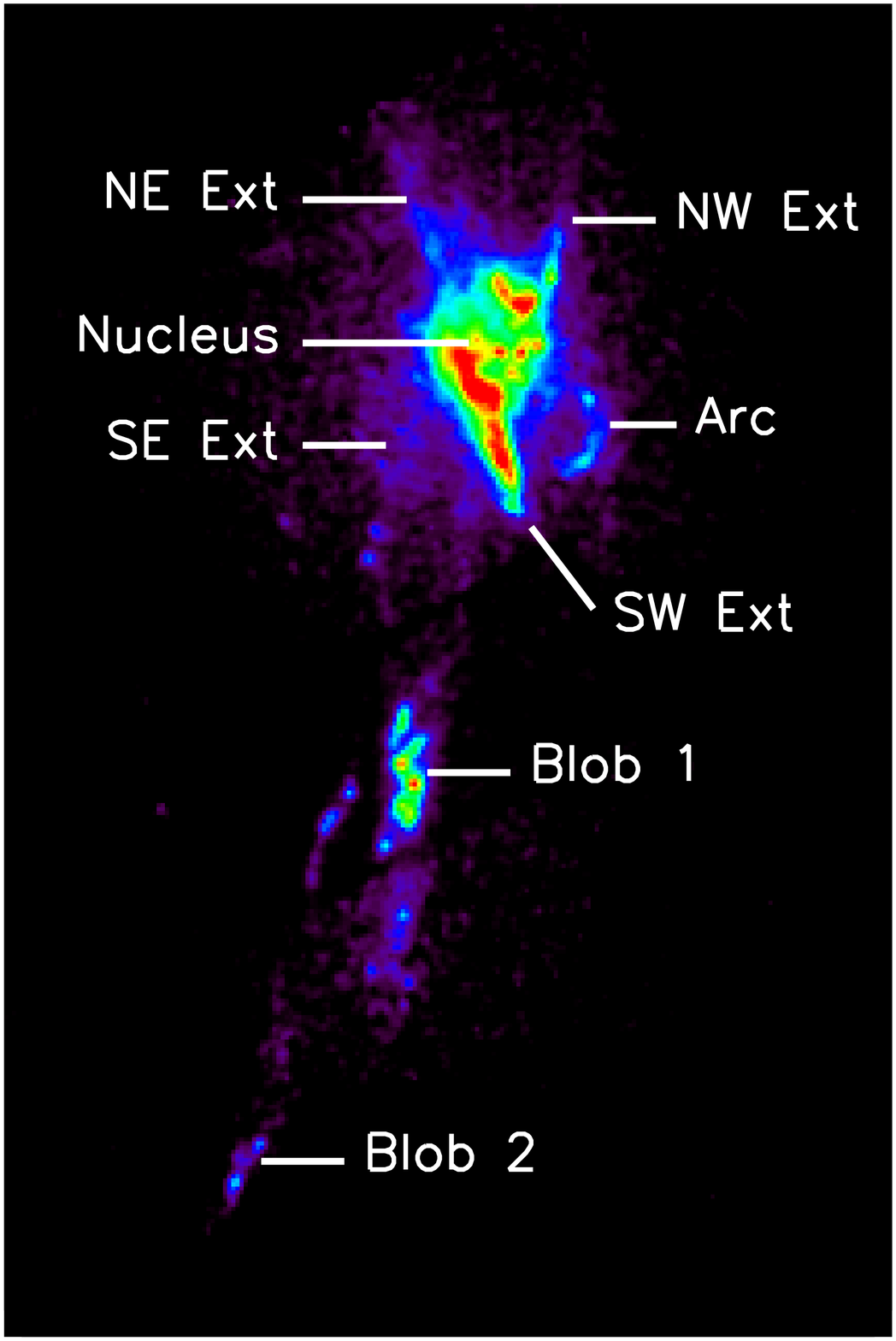} \\
\end{tabular}
\label{multiband}
\caption{MMTF H$\alpha$ and [N~II] $\lambda$6583 and {\em HST}/SBC FUV
  images of Abell~1795. The three upper panels show maps of the
  surface brightness, $\Sigma$, in units of ergs s$^{-1}$ cm$^{-2}$
  arcsec$^{-2}$. The lower left and center panels are [N~II]/H$\alpha$
  and FUV/H$\alpha$ ratio maps, respectively. In the latter image, the
  red shade represents regions where there is very little or no
  H$\alpha$ coincident with the FUV emission. The bottom right panel
  outlines the terminology we use in this paper, specifically in Table
  1, for the different regions. The ``central region'' consists of all
  emission north of (and including) the SW extension. In all panels,
  the 10 kpc scale corresponds to 7.9$^{\prime\prime}$, and the black
  crosshair represents the centroid of optical emission, at $\alpha=$207.2188$^{\circ}$, $\delta=$26.5929$^{\circ}$.}
\end{figure*}


\section{Observations and Data Reduction}

\subsection{H$\alpha$ -- Maryland-Magellan Tunable Filter}

MMTF has a very narrow bandpass ($\sim$5--12\AA) which can be tuned to
any wavelength over $\sim$5000-9200\AA~(Veilleux \etal 2009). Coupled
with the exquisite image quality of Magellan, this instrument is ideal
for detecting emission-line filaments in distant clusters. During
April 2008, we observed Abell~1795 for a total of 60 minutes each at
$\lambda_{H\alpha}=6972.8$ \AA, $\lambda_{[NII]}=6994.7$ \AA\ and
$\lambda_{continuum}=7044$ \AA. These data were reduced using the MMTF
data reduction
pipeline\footnote{http://www.astro.umd.edu/$\sim$veilleux/mmtf/datared.html}.
The continuum image was then PSF and intensity matched to the
narrow-band images to allow for careful continuum subtraction.

\subsection{Far UV -- Hubble Space Telescope ACS/SBC}

FUV imaging was acquired using the ACS SBC on the {\em HST} in the
F140LP bandpass, with a total exposure time of 1197 seconds. With the
MMTF data already in hand, we were able to choose two different
pointings of the 35$\arcsec\times$35$\arcsec$ field of view to allow
full coverage of the SE filament. Exposures with multiple filters are
required to properly remove the known SBC red leak, which may be
substantial in the central region due to the fact that the underlying
galaxy is very luminous and red. However, we were unable to obtain
complementary exposures in a redder SBC band for the filaments, due to
scheduling constraints, and thus we proceed without removal of the
offending light. The central UV fluxes have an associated error of
$\sim$~2.5\% due to the underlying red galaxy (based on a preliminary
analysis of additional FUV data of the central galaxy obtained by W.\
Sparks and collaborators).

\section{Results}

\begin{figure}[t]
\centering
\includegraphics[width=0.48\textwidth]{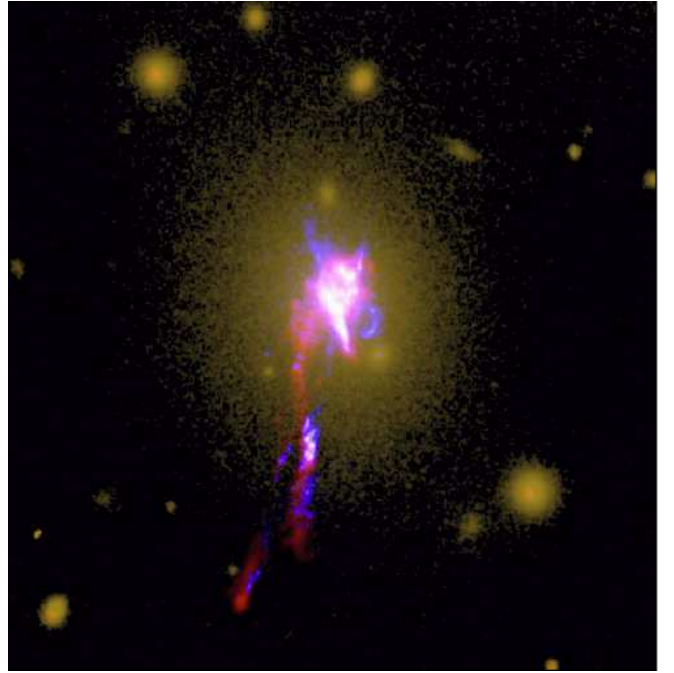}
\caption{Color composite image of Abell~1795. The three colors
  represent the red continuum (yellow), H$\alpha$ (red) and far UV
  (blue). The long SE filament is 
  resolved into two intertwined filaments in H$\alpha$ and FUV. 
}
\label{colorcomp}
\end{figure}

\begin{figure}[htb]
\centering
\includegraphics[width=0.45\textwidth]{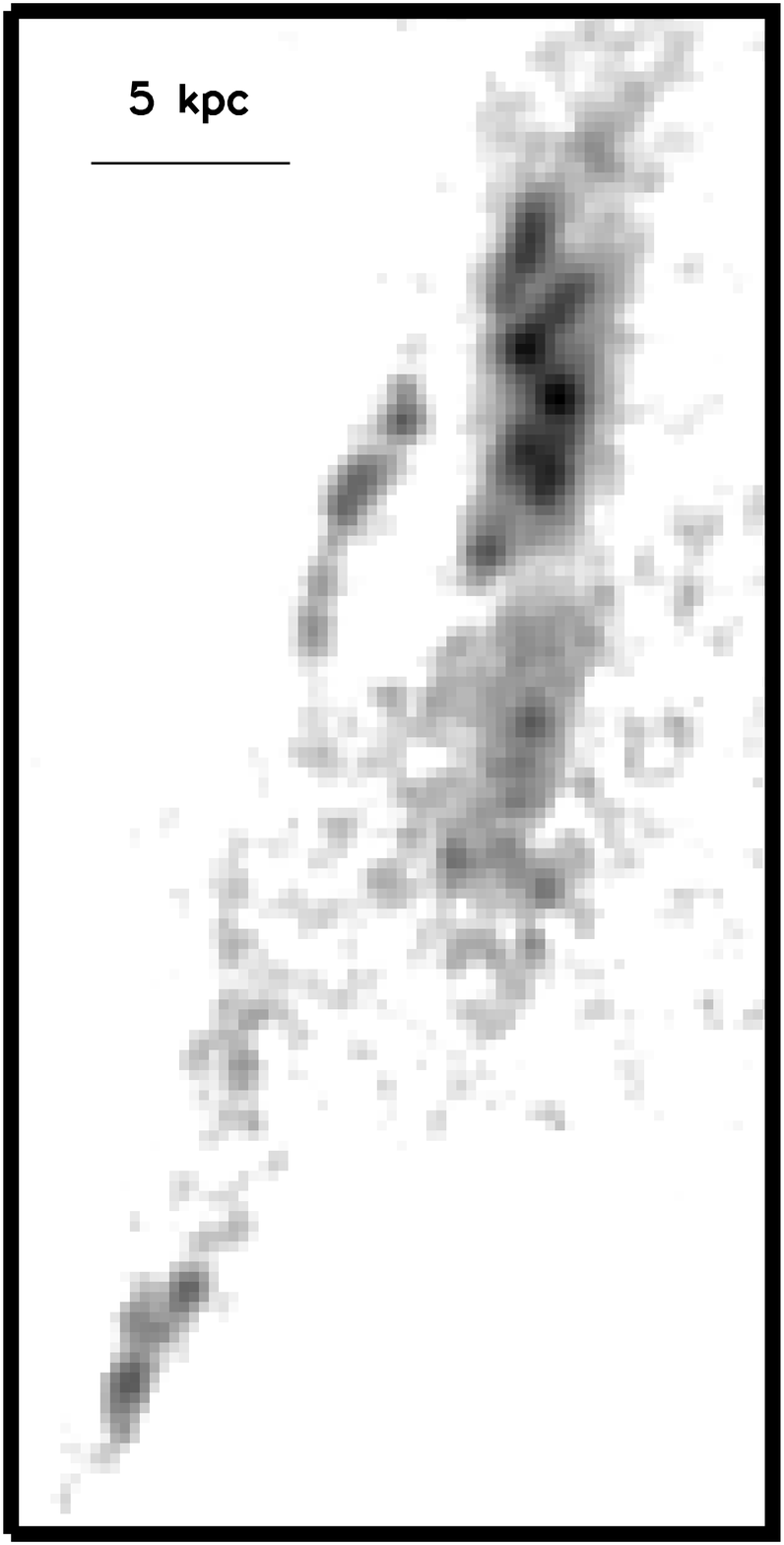}
\caption{Close-up view of the criss-crossing filaments in the FUV
  showing both of the bright blobs. The brightest portions of the
  filaments break up into young UV-bright super star clusters.}
\label{uvfil}
\end{figure}

Figures \ref{multiband} -- \ref{uvfil} show the newly acquired MMTF
and {\em HST} data on Abell~1795. The most striking result is that the
``SE filament'', which has long been known (Cowie \etal 1983) is, in
fact, a pair of thin, intertwined filaments in H$\alpha$. These
filaments are $\sim$ 42$\arcsec$ and 35$\arcsec$ (52.9 and 44.1 kpc)
in length, and their widths in H$\alpha$ are unresolved ($<0\farcs 7
\sim 1$ kpc). The discovery of thin strands in the SE filament is
reminiscent of magnetic field lines. A stronger than average magnetic
field in the ICM could prevent the filament material from evaporating
due to thermal conduction. The BCG in Abell~1795 is a bright
double-jet radio-loud cD galaxy (4C~26.42) with therefore strong
extended magnetic field (Ge \& Owen 1993), but radio emission is not
detected beyond $\pm$6 kpc from the nucleus. We return to this issue
in Section 4.

In the central region of Abell~1795, the H$\alpha$ emission forms a
ring, with a cavity slightly northeast of the BCG center (Figure
\ref{multiband}). The morphology is very similar to that seen in the
X-ray (Fabian \etal 2001). The cavity surrounds one of the radio jets
emanating from the AGN, suggesting that it was created by recent AGN
activity (Crawford \etal 2005). The very sharp extension directly to
the southwest of the central region is coincident with the
counter-jet. The jet appears to be efficiently heating the gas in this
region (van Breugel \etal 1984). In general, the H$\alpha$ emission
spatially correlates well with the X-ray emission in both the central
regions and on larger scales to within the spatial resolution of the
X-ray data.

The upper middle panel of Figure \ref{multiband} reveals the
distribution of [N~II] in Abell~1795. All of the features that are
seen in H$\alpha$ are also visible in [N~II], including both
filaments, the short SW jet, the nucleus and the cavity northeast of
the nucleus. The ratio of [N~II]/H$\alpha$, a measure of the relative
importance of heating and ionization, is high and fairly uniform
throughout the filaments ($\sim 0.35$--$0.55$, see Table 1 and the
lower left panel of Figure \ref{multiband}); this elevated ratio
confirms earlier results (Crawford \etal 2005 and references therein)
and seems inconsistent with photoionization by hot
stars. [N~II]/H$\alpha$ is even higher in the nucleus, approaching
unity, likely due to heating by the AGN, and drops northeast of the
nucleus, where the jet has apparently cleared out a cavity. Given the
MMTF bandwidth ($\sim$10\AA), we note that the [N~II] image will also
contain any H$\alpha$ with relative velocity $>$785 km s$^{-1}$, and
vice versa. However, since the typical velocity widths are $\sim$300
km s$^{-1}$ in the filaments (e.g.\ Crawford \etal 2005), we do not
expect interline contamination to be important.

In the FUV (Figure \ref{multiband}, upper right panel, and Figure
\ref{uvfil}), both strands are visible and, with the added resolution
of {\em HST}, are resolved in some regions into chains of bright,
compact sources -- the sites of recent star formation (see also
Crawford \etal 2005).  The two brightest blobs seen in the filaments
at H$\alpha$ (blobs \#1 and \#2 in the nomenclature of Figure
\ref{multiband}) break up into FUV point-source and diffuse emission,
but spatial offsets of $\sim$1--2 kpc are visible between the
H$\alpha$ and FUV emission centroids. This shift is particularly
obvious in the FUV/H$\alpha$ map in the lower middle panel of Figure
\ref{multiband}. We do not believe this relative offset is due to
errors in astrometry, since it was also noted in the earlier data of
Crawford \etal (2005).

\begin{table*}[htb]
\centering
\caption{Luminosities, flux ratios and inferred star formation rates in 
  the central region of Abell~1795.}
\label{lumtable}
\begin{tabular} {lcccccrr}
\\
\hline\hline\\
Region & L$_{H\alpha}$ & L$_{[NII]}$ & L$_{UV}$ & [N~II]/H$\alpha$ & UV/H$\alpha$ &  SFR$_{H\alpha}^{(a)}$ & SFR$_{UV}^{(a)}$\\
$$ & [10$^{7}$ L$_{\odot}$] & [10$^{7}$ L$_{\odot}$] & [10$^{8}$ L$_{\odot}$] & $$ & $$ & [M$_{\odot}$/yr] & [M$_{\odot}$/yr]\\\\
\hline\\
Total            & 5.49 & 1.97 & 8.07 & 0.36  &  14.7 & $<$1.67 & $<$2.66\\
\\
~~SE filaments   & 0.72 & 0.26 & 1.47 & 0.36 &  20.3 & 0.22 & 0.48\\
~~~~ Blob 1      & 0.25 & 0.09 & 0.70 & 0.38 &  28.4 & 0.08 & 0.23\\
~~~~ Blob 2      & 0.16 & 0.09 & 0.16 & 0.53 &  10.0 & 0.05 & 0.05\\
\\
~~Central region & 4.77 & 1.72 & 6.61 & 0.36 &  13.8 & $<$1.45 & $<$2.18\\
~~~~Nucleus      & 0.12 & 0.10 & 0.06 & 0.81 &   4.4 & $<$0.04 & $<$0.02\\
~~~~NE extension & 0.05 & 0.00 & 0.36 & 0.00 &  73.0 & 0.02 & 0.12\\
~~~~NW extension & 0.05 & 0.00 & 0.16 & 0.00 &  33.3 & 0.02 & 0.05\\
~~~~SE extension & 0.50 & 0.13 & 0.44 & 0.26 &   8.9 & 0.15 & 0.15\\
~~~~SW extension & 1.00 & 0.48 & 1.85 & 0.48 &  18.5 & 0.30 & 0.61\\
~~~~Arc          & 0.03 & 0.00 & 0.28 & 0.00 & 102.8 & 0.01 & 0.09\\
\\
\hline
\\
\end{tabular}
\tablenotetext{(a)}{ Star formation rates derived using the
prescriptions in Kennicutt (1998) assuming all of the H$\alpha$ and UV
emission, including that from the nucleus, is due to star formation;
the value in the nucleus is therefore an upper limit.}
\newline
\end{table*}

Most of the bright H$\alpha$ features in the central region emit
strong FUV, such as the nucleus, the ring and the SW
extension. However, there are features in the central region that are
not common between the FUV and H$\alpha$ data. In the FUV map, there
are two northern extensions to the east and west (the NE and NW
``extensions'' in Figure \ref{multiband}), as well as a curved
extension southwest of the central region (the ``arc''). The emission
in these regions is lumpy but largely unresolved into points
sources. In H$\alpha$, there is a very bright and thick extension at
the base of the SE filament which is not nearly as bright in the
FUV. This region and other anomalous features are easily identified in
the FUV/H$\alpha$ map. (More details will be given in an upcoming
paper where images of the central region of Abell~1795 in all three
FUV bands of ACS/SBC are considered. This multiband analysis is beyond
the scope of the present {\em Letter}.)


\section{Discussion}

\subsection{Origin and Power Source of Filaments}

A number of very specific scenarios were put forward by Fabian \etal
(2001) to explain the origin of the SE filament, taking into account
that 4C~26.42 is in motion within the gravitational potential of the
Abell~1795 cluster (Oegerle \& Hill 2001; see also discussion in
Crawford \etal 2005 and Rodriguez-Martinez \etal 2006): (a) a cooling
wake, produced by a cooling flow occurring around the moving cD galaxy
(as in NGC~5044; David \etal 1994), (b) a ``condensation trail''
produced by the ram pressure of the radio source passing through the
multiphase medium of the ICM and ISM of the host galaxy, (c)
evaporation of cold gas ram-pressure stripped from the cD galaxy, and
(d) an accretion wake produced by the gravitational focussing effects
of the moving cD galaxy on the ICM (e.g. Sakelliou \etal 1996).

Scenario (d) requires that the accreting ICM be gravitationally
focused by the cD galaxy and cool into a wake. However, the relatively
large sound speed of the hot gas makes it unlikely that this process
alone can account for the presence of the long, thin, X-ray
filament. In scenario (c), the material making up the filaments is ISM
stripped from the cD galaxy. The large mass of hot gas in the
filaments, $\sim 5\times 10^9$ M$_\odot$ (Fabian \etal 2001), and thin
geometry of the filaments are hard to explain in this scenario.
However, this does not exclude the possibility that {\em some} of the
gas outside the cD galaxy is produced in this way (e.g.\ the broad
base of the H$\alpha$ filaments).  Scenario (b) also presents some
problems. As argued by Fabian \etal (2001), it seems unlikely that the
cD galaxy plunging through the ICM would result in cooling, rather
than heating, of the multiphase medium. Moreover we see no obvious
connection in the data between the radio jets of 4C~6.42 and the long
filaments.

By process of elimination, these arguments seem to favor scenario (a).
The very thin geometry of the H$\alpha$ filaments points to highly
non-linear runaway cooling in this region. Potentially contributing
heating/ionization sources in the filaments include (i) the central
AGN, (ii) hot young stellar population outside the cD galaxy, (iii)
X-rays from the ICM itself, (iv) heat conduction from the ICM to the
colder filaments, (v) shocks and turbulent mixing layers, and (vi)
collisional heating by cosmic rays. Arguments based on the AGN
energetics and the line ratios in and out of the central region
suggest that the AGN is not strongly influencing the ionization of the
material much beyond the central region ($R \la$ 6 kpc). The long,
thin geometry of the filaments and their quiescent velocity field
(Crawford \etal 2005) seem to rule out ionization by shocks and/or
turbulent mixing layers, while the relative weakness of
high-ionization lines suggests that ionization by the X-ray ICM is
also a small contributor. This leaves scenarios (ii), (iv), and (vi)
as the most plausible explanations for the dominant heating mechanism
in the filaments.

\begin{figure}[htb]
\centering
\includegraphics[width=0.5\textwidth]{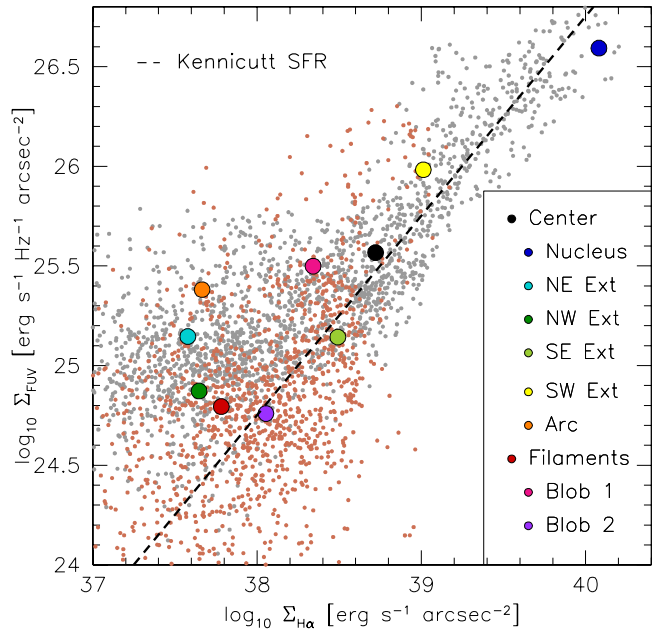}
\caption{FUV surface brightness, $\Sigma_{FUV}$, versus H$\alpha$
  surface brightness, $\Sigma_{H\alpha}$, in Abell~1795.  The dashed
  line represents the expected relation between the global
  $\Sigma_{H\alpha}$ and $\Sigma_{FUV}$ for star-forming galaxies
  (Kennicutt 1998). The larger symbols show the average surface
  brightness over the brighter features defined in Figure
  \ref{multiband}. The background distribution of points shows the
  2$\times$2 pixel-by-pixel surface brightness measurements for the
  central region (grey) and filaments (light red). }
\label{sfr}
\end{figure}

The presence of FUV point sources in the filaments (Figure
\ref{uvfil}) lends support to scenario (ii), but is there enough star
formation to account for the observed H$\alpha$ emission? To try to
answer this question, we compare the FUV and H$\alpha$
emission. Figure \ref{sfr} shows the average H$\alpha$ and FUV surface
brightnesses of the brighter features defined in Figure
\ref{multiband}. The H$\alpha$ and FUV luminosities of these features
are listed in Table 1.  Assuming that the FUV and H$\alpha$ star
formation rate prescriptions of Kennicutt (1998) derived from the
global properties of star-forming galaxies also apply individually to
these features (we return to this assumption below), we find that the
H$\alpha$ and FUV surface brightnesses and luminosities of these
features imply star formation rate surface densities and star
formation rates which are generally consistent to within a factor of
$\sim$2 of each other. (The only exceptions are the arc and the NE and
NW extensions, which are underluminous in H$\alpha$; this could be due
to a lack of gas in these regions or other forms of continuum emission
process. A detailed analysis of the central region using all three FUV
bands of ACS/SBC (PI: W.\ Sparks) is in preparation.) Under this
assumption, the total H$\alpha$ and UV-determined star formation rates
are $\la$1.7 -- 2.7 M$_{\odot}$ yr$^{-1}$, with $\sim$0.2 -- 0.5
M$_{\odot}$ yr$^{-1}$ ($\sim$15 -- 25\%) contained in the filaments
(Table 1).  These numbers are lower than previous measurements of the
total, integrated star formation rate of $\sim$ 5-20 M$_{\odot}$
yr$^{-1}$ (Smith \etal 1997; Mittaz \etal 2001) and the estimate of
$\la$ 1 M$_{\odot}$ yr$^{-1}$ of star formation in the filament
(Crawford \etal 2005). These are all $\sim$2 orders of magnitude
smaller than the predicted {\em Chandra-} and {\em XMM}-derived
integrated mass deposition rates from the inner ICM (Ettori et
al. 2002; Peterson et al. 2003).

The integrated H$\alpha$ and FUV quantities discussed so far do not
tell the full story: Large spatial variations of the FUV/H$\alpha$
ratio are seen on smaller scale (see lower middle panel of Figure
\ref{multiband} and pixel-to-pixel surface brightness measurements of
Figure \ref{sfr}).  These spatial variations may be due to a number of
effects, including differential dust extinction, variations in star
formation history (SFH: age of burst, decay time scale) or initial
mass function (IMF), complex geometry of the gas relative to the
ionizing stars, and contributions from non-stellar processes to the
FUV and H$\alpha$ emission.  Extinction corrections would boost the
FUV fluxes relative to H$\alpha$ and therefore could account for
regions with anomalously small FUV/H$\alpha$ ratios. Published data on
the filaments suggest relatively modest extinctions, however (e.g.\
Crawford \etal 2005). Variations in the SFH or IMF change the relative
importance of ionizing and non-ionizing stars and may account for
variations in both directions of the FUV/H$\alpha$ flux ratio. Our
data do not provide strong constraints on these parameters.
Geometrical effects are undoubtedly important in some regions,
particularly in blobs \#1 and \#2, where spatial offsets of $\sim$1--2
kpc are visible between the H$\alpha$ and FUV emission centroids. In
these blobs, the Kennicutt (1998) prescriptions may severely
underestimate the number of FUV-bright stars needed to account for the
H$\alpha$ emission; processes other than photoionization by hot stars
appear needed to account for the observed H$\alpha$ at these locations
(the recombination time scale is at least an order of magnitude
shorter than the dynamical time scale to move $\sim$1--2 kpc, unless
the density of the H$\alpha$ filaments is much less than $\sim$1
cm$^{-3}$).

As mentioned in Section 3, additional heating sources also appear
needed to explain the unusually strong low-ionization lines detected
in the blobs and in between them (e.g.  Hu \etal 1985; Crawford \etal
2005; Figure \ref{multiband}).  Two heating processes remain viable:
(iv) heat conduction from the ICM to the colder filaments and (vi)
collisional heating by cosmic rays. The relative importance of these
processes critically depends on the strength of the magnetic field in
the filaments. Strong magnetic fields could shield the cooling ICM gas
from re-heating by conduction with the hot ICM (scenario (iv)),
creating long tubes of cool, dense gas.  Runaway star formation would
take place along these magnetic field lines, where the gas is cooling
and condensing to high enough density to become Jeans unstable. The
long, thin geometry of the H$\alpha$ SE filaments and detection of
embedded FUV-bright stellar clusters in these filaments are consistent
with this picture.

Ferland \etal (2008, 2009) have recently examined the question of the
importance of heating by energetic particles or dissipative
magnetohydrodynamic (MHD) waves in the central nebulae of massive
clusters. The energetic particles contributing to the heating of the
filaments in this scenario may either be produced in-situ by MHD
processes or conducted in from the surrounding intracluster medium.
Given our previous discussion, we speculate that the optical emission
in the SE filaments of Abell~1795 may naturally be explained by these
heating processes. The magnetic field in the filaments may represent
residual magnetic field originally associated with the radio galaxy
but now entrained in the ICM flow. If this is the case, the
crisscrossing geometry of the SE filaments may reflect precession of
the radio jets or the orbital motion of the cD galaxy in the cluster
potential. 

Finally, we end with a cautionary note: Our data do not provide
quantitative constraints on the strength of the purported ICM magnetic
field. The possibility of runaway star formation unaided by magnetic
field cannot be ruled out. In fact, the filaments seen in Abell~1795
shares a morphological resemblance with the narrow cold
streams seen feeding galaxies in recent high-resolution numerical
simulations of the early universe (e.g.\ Ceverino \etal 2009).

\section{Concluding Remarks}

Using deep, high-resolution H$\alpha$, [N~II] $\lambda$6584, and FUV
data, we have discovered that the SE filament in Abell~1795 is in fact
two intertwined filaments of ionized hydrogen. The most plausible
origin for these filaments is a wake of cooling ICM behind the moving
cD galaxy in Abell~1795. The narrowness of the strands suggest highly
non-linear runaway cooling of the ICM. Their tangled morphology
suggests that the infalling gas may be interacting with enhanced
magnetic fields, allowing for less efficient energy conduction and
faster cooling in this region. We observe knots of UV-bright point
sources along these filaments, indicating star formation at a rate of
$\sim$0.5 M$_\odot$~yr$^{-1}$ in the filaments. The large spatial
variations of the FUV/H$\alpha$ ratio and enhanced low ionization
lines suggest that O-star photoionization is not the sole source of
heating of this gas; collisional heating by energetic particles is
another likely contributor. A deeper understanding of the origin of
the filaments of Abell~1795 will require detailed MHD modeling and
observations to constrain the purported magnetic field, both of which
are beyond the scope of the present {\em Letter}.  It is also not
clear yet whether this scenario applies to cooling flow clusters in
general. We plan to address this issue in upcoming papers using a
representative set of massive clusters.

\acknowledgements Support for this work was provided to M.M.\ and
S.V.\ by NSF through contract AST 0606932 and by NASA through contract
HST GO-1198001A.~~S.V. also acknowledges support from a Senior Award
from the Alexander von Humboldt Foundation and thanks the host
institution, MPE Garching, where this paper was written. We thank
D.S.N.\ Rupke, C.S.\ Reynolds, E.\ Ostriker, H.\ Netzer, and M.C.\
Miller for useful discussions, and are particularly grateful to W.\
Sparks for letting us examine F150LP and F165LP images of the central
region of Abell~1795 obtained under HST/PID \#11681.

\end{document}